\documentclass[aps,prl,twocolumn,showpacs,superscriptaddress,floatfix,nofootinbib]{revtex4-1}
\usepackage{graphicx,graphics}
\usepackage{dcolumn}
\usepackage{amsmath,amssymb,amsfonts}
\usepackage{latexsym,verbatim}
\usepackage{bm}
\usepackage{color}
\usepackage[breaklinks=true,colorlinks,citecolor=blue,linkcolor=blue,urlcolor=blue]{hyperref}
\usepackage{tabularx}

\def\be{\begin{eqnarray}}
\def\ee{\end{eqnarray}}

\begin{document}
\title{Suppressing backscattering of helical edge modes with a spin bath}

\author{Andrey A. Bagrov}
\email{a.bagrov@science.ru.nl}
\affiliation{Institute for Molecules and Materials, Radboud University, Heijndaalseweg 135, 6525 AJ, Nijmegen, The Netherlands}
\author{Francisco Guinea}
\email{paco.guinea@gmail.com}
\affiliation{IMDEA-Nanoscience, Calle de Faraday 9, E-28049 Madrid, Spain}
\affiliation{School of Physics and Astronomy, University of Manchester, Manchester M13 9PY}

\author{Mikhail I. Katsnelson}
\email{m.katsnelson@science.ru.nl}
\affiliation{Institute for Molecules and Materials, Radboud University, Heijndaalseweg 135, 6525 AJ, Nijmegen, The Netherlands}

\begin{abstract}
In this paper, we address the question of stability of protected chiral modes (e.g., helical edge states at the boundary of two-dimensional topological insulators) upon interactions with the external bath. Namely, we study how backscattering amplitude changes when different interaction channels between the system and the environment are present. Depending on the relative strength of the Coulomb and spin-spin channels, we discover three different possible regimes. While the Coulomb interaction on its own naturally amplifies the backscattering and destroys the protection of chiral modes, and the spin-spin channel marginally suppresses backscattering, their interplay can make the backscattering process strictly irrelevant, opening the possibility to use the external spin bath as a stabilizer that alleviates destructive effects and restores the chirality protection.
\end{abstract}

\maketitle

Topological insulators (TI) are characterized by existence of protected helical edge states, - one-dimensional chiral modes at the edges of two-dimensional TI, and two-dimensional massless Dirac fermions at the surfaces of three-dimensional TI \cite{TI1,TI2,TI3,TI4,TI5,TI6,TI7}. This is a manifestation of a very general ``bulk-edge correspondence'' principle \cite{Hatsugai1,Hatsugai2,Prodan, Slager} which is probably one of the brightest applications of topological and geometrical concepts in condensed matter physics. Importantly, topological protection of the edge states is not absolute: they can be broken by spin-dependent scattering mechanisms such as scattering on magnetic impurities \cite{TI6, Mirlin, Burmistrov1, Burmistrov2} or electron-electron interactions \cite{Glazman}. These factors result in the backscattering and destruction of the helical modes, due to the intimate relation between their propagation direction and direction of spins: if one flips the spin, one reverses the momentum. This effect has been considered from many perspectives, and a variety of its possible physical consequences on the transport and spectral properties of helical channels have been studied (see e.g. \cite{Hohenadler1, Hohenadler2}).

Because of the importance of practical implications of chiral edge modes, it is interesting to think of possible ways to reduce (or even eliminate) backscattering and make the edge modes more stable.
To achieve this goal, we suggest to couple the channel to a spin environment. While environment consisting of static spin degrees of freedom acts as a set of magnetic impurities that induce and amplify backscattering \cite{Klinovaja1, Klinovaja2}, the physics of fast itinerant spins can be very different, as known in the theory of magnetice resonance \cite{Book1, Book2}. According to the popular decoherence program in quantum physics \cite{P1,P2} (for the recent critical discussion of this program, see \cite{P3}), instant interactions between the environment and the channel can be thought of as effective von Neumann projective measurements that tend to make the spins classical via the ``orthogonality catastrophe'': the environment degrees of freedom get entangled with spin-up and spin-down states of the system, and the small overlap of corresponding wavefunctions suppresses the amplitude of spin-flip processes \cite{Leggett1,Leggett2,Stamp}. A related model, where dissipation induces decoherence in a Luttinger liquid, has been studied in \cite{CSG06}. For the edge modes, this would mean stabilization of the states with definite momenta; in terminology of Zurek \cite{P1}, they appear to be ``pointer states'' robust with respect to the interaction with the environment. This situation looks unusual: in most of the cases the interactions between the central system and the environment are much stronger dependent on the coordinates than on the momenta, which tends to stabilize the states with definite coordinates, i.e. localized in real space, rather than the states with definite momenta \cite{P2}. Here we provide a formal analysis of the effect the environment has on the backscattering of helical states, using the renormalization group approach similar to the one used in \cite{Leggett2,RG1,RG2,RG3,RG4}. It turns out that, depending on the ratio of the exchange and direct interactions, the environment can both suppress and enhance the backscattering.

We start with the following one-dimensional $s-d$ model, which, albeit simple, captures all the relevant aspects of more complicated and peculiar systems:
\begin{gather}\label{eq:Hamiltonian}
  {\cal H} = \sum\limits_kc^\dagger(k) H^c(k) c(k) + \sum\limits_{k;i=1,2} d_i^\dagger(k)H^d_i(k) d_i(k)- \\
  J \sum\limits_q \left(\sum\limits_k c^\dagger(k) \vec{\sigma} c(k+q) \right)
  \left(\sum\limits_{p;i=1,2} d^\dagger_i(p)\vec{\sigma} d_i(p-q)\right), \nonumber
\end{gather}
where $\vec{\sigma}$ are the Pauli matrices, $k$ is the one-dimensional spatial momentum, and the standard notation is used:
\begin{equation}
 \sum\limits_k = \int\limits_{-\pi/a}^{\pi/a} \frac{a d k}{2\pi},
\end{equation}
where $a$ is the lattice constant.
Here $c(k)$ and $d_{1,2}(k)$ are the chiral edge modes of topological insulator and the environment degrees of freedom, respectively:
\begin{gather}
c(k) = (c^{\uparrow}(k), c^{\downarrow}(k)),\\
d_i(k) = (d_i^{\uparrow}(k), d_i^{\downarrow}(k)), \nonumber
\end{gather}
and the Hamiltonians of each sector are given by
\begin{gather}\label{eq:KineticC}
H^c(k)=\left( \begin{array}{cc} \hbar v_F k & h_0 \\ h_0 & -\hbar v_F k \end{array} \right), \\
H^d_{1,2}(k)=\left( \begin{array}{cc} \pm \hbar c k & 0 \\ 0 & \pm \hbar c k \end{array} \right). \label{eq:KineticD}
\end{gather}
Since there is no preferred helicity in the environment, we take into account both right-moving ($i=1$) and left-moving ($i=2$) particles, and represent them for simplicity as two independent fermionic flavors.
The bare backscattering is introduced via the off-diagonal term $h_0$ of the edge modes Hamiltonian. A concrete mechanism that induces backscattering is not important for our considerations.

In what follows, we will analyze how the parameter controlling backscattering changes due to the interactions with the spin environment, relying upon perturbative renormalization group approach \cite{Leggett2,RG1,RG2,RG3,RG4}. As it will be evident, other interaction channels will be induced on top of the isotropic spin-spin interaction introduced in the Hamiltonian \eqref{eq:Hamiltonian}, and it turns out to be convenient to include them into the original Hamiltonian as a generalized vertex:
\begin{gather}
{\cal H}_{int} = \Gamma^{(i)}_{\alpha\beta\gamma\delta} \sum\limits_{q,p,k} c_{\alpha}^{\dagger}(k)  c_{\beta}(k+q)
  d_{i,\gamma}^{\dagger}(p) d_{i,\delta}(p-q), \label{eq:GenVertex}\\
 \label{eq:IntVertex} \Gamma^{(i)}_{\alpha\beta\gamma\delta} = J^{(i)}_{00} \mathbb{I}_{\alpha\beta} \otimes \mathbb{I}_{\gamma\delta} + J^{(i)}_{zz} \sigma^z_{\alpha\beta} \otimes \sigma^z_{\gamma\delta} + \\ J^{(i)} \left(\sigma^x_{\alpha\beta} \otimes \sigma^x_{\gamma\delta}  + \sigma^y_{\alpha\beta} \otimes \sigma^y_{\gamma\delta}\right) + \nonumber \\ J^{(i)}_{0z} \mathbb{I}_{\alpha\beta} \otimes \sigma^z_{\gamma\delta} + J^{(i)}_{z0} \sigma^z_{\alpha\beta} \otimes \mathbb{I}_{\gamma\delta}, \nonumber
\end{gather}
where we also added the Coulomb channel $J_{00}$, the spin-charge channels $J_{0z}$ and $J_{z0}$, and the possible anisotropy between $Z$ and $XY$ spin couplings. This reduces to the isotropic spin interaction of \eqref{eq:Hamiltonian} if
\begin{equation}
J^{(i)}_{00}=J^{(i)}_{0z}=J^{(i)}_{z0}=0,\,\,\,\,J^{(i)}_{zz}=J^{(i)}=J.
\end{equation}
To make the notations more handy and reduce the number of indices,
hereinafter we denote the coupling constants $J^{(1)}$ as plain $J$, and $J^{(2)}$ as $\tilde{J}$.

As we elaborated in the introduction, we expect the spin-spin interactions between the edge of the topological insulator and the bath to make pointer states of the system to be states with well-defined spin, and thus stabilize the helical modes. In terms of the renormalization group flow for the model \eqref{eq:Hamiltonian}, \eqref{eq:GenVertex}, it means that the mode-mixing parameter $h$ is expected to become irrelevant in the infrared.

\begin{figure}[t]
\begin{center}
\includegraphics[width=1\columnwidth]{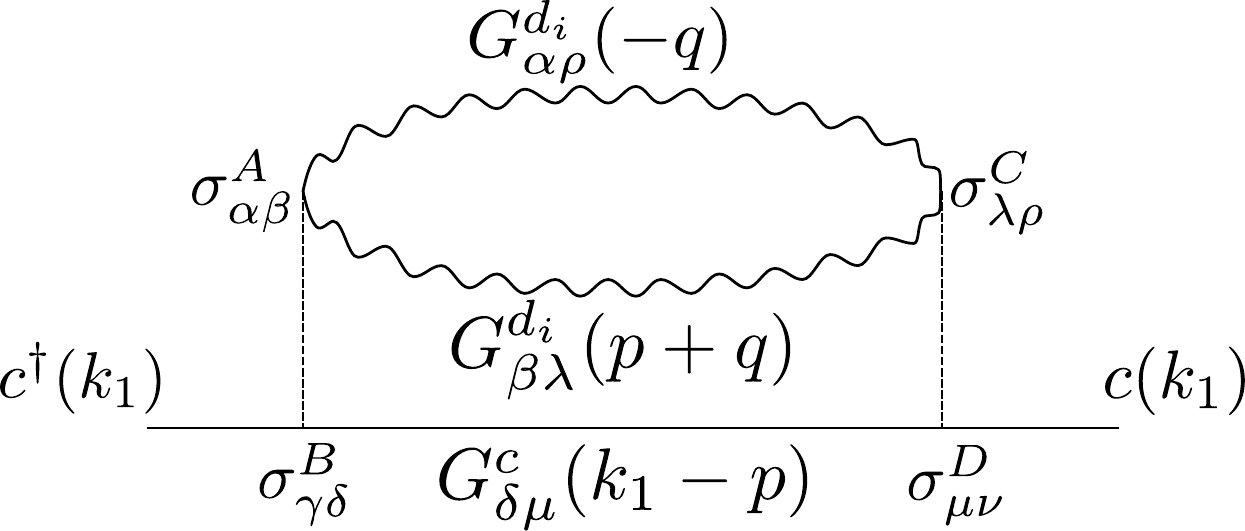}
\end{center}
\caption{Self-energy correction to the helical edge modes. Wavy lines denote the propagators of the environment modes. Latin letters stand for $x,y,z$, and the Greek ones denote the spin indices. Sum over all combinations of $A,B,C,D$ allowed by the structure of vertex \eqref{eq:IntVertex} has to be taken. \label{fig:SelfEnergy}}
\end{figure}

The leading order quantum correction to $h$ is given by off-diagonal part of the two-loop self-energy diagram shown in Fig. \ref{fig:SelfEnergy} (from now on all calculations will be conducted for the Matsubara Green's functions):
\begin{gather}
G_c^{-1}(i\omega, k) = G^{-1}_{c,0}(i\omega, k) - \Sigma(i\omega, k), \\
h(i\omega, k) = h_0 + \Sigma_{01}(i\omega, k), \nonumber
\end{gather}
where the bare Green's function of edge fermions is related to their Hamiltonian \eqref{eq:KineticC} as
\begin{equation}
G^{-1}_{c,0}(i\omega, k) = i\omega \cdot \mathbb{I} - H^c(k).
\end{equation}
The polarization loop is given by a simple integral:
\begin{gather}
\Pi_{1,2}^{AC}(p) =
\int\limits_{-\pi/a}^{\pi/a} \frac{a d q}{2\pi}
\int\limits_{-\infty}^\infty \frac{d\omega_q}{2\pi} \mbox{Tr} \left[\sigma^A G^{d_{1,2}}(i\omega_p+i\omega_q,p+q)\cdot \right. \\ \left. \sigma^C G^{d_{1,2}}(-i\omega_q,-q)\right] \nonumber = \frac{a p}{\pi(\mp i\omega_p + \hbar c p)} \delta^{AC}
\end{gather}
To obtain the self-energy correction, we need to sum over all possible combinations of $A,B,C,D$ indices in Fig. \ref{fig:SelfEnergy} that give non-trivial contributions, as well as over the two flavors of the environment modes. The resulting expression at zero external momentum is
\begin{gather}
\Sigma_{01}(0,0) = -\int\limits_{-\pi/a}^{\pi/a} \frac{a d q}{2\pi}\int\limits_{-\infty}^\infty \frac{d\omega_p}{2\pi}\frac{ah q}{(h^2+\omega_q^2 +\hbar^2 v_F^2 q^2)} \cdot \nonumber \\ \left(\frac{\alpha(J)}{i\pi \omega_q -\pi \hbar c q} - \frac{\alpha(\tilde{J})}{i\pi \omega_q +\pi \hbar c q} \right),
\end{gather}
where we introduced
\begin{gather}
 \alpha(J)=J^2_{00} + J^2_{0z} - J^2_{z0} - J^2_{zz}.
\end{gather}
Although there is a natural ultraviolet (UV) cut-off given by the lattice constant $a$, it is convenient to formally consider the momentum integral over the second loop as logarithmically divergent in the $a \rightarrow 0$ limit, as it allows to
extract the leading scaling that defines the renormalization group flow. Evaluating the integral over frequencies via residues, and then expanding the integrand around $|q| \rightarrow \infty$, we obtain the correction to backscattering amplitude from a thin momentum shell $|q| \in \left[ \Lambda, \Lambda + d\Lambda \right]$:
\begin{gather}
h(\Lambda + d\Lambda) = h(\Lambda) + \delta \Sigma_{01} = h(\Lambda) - \\
\frac{2}{4\pi^2}\int\limits_\Lambda^{\Lambda+d\Lambda}\frac{a^2h(q)[\alpha(J)+\alpha(\tilde{J})]d q}{\sqrt{h(q)^2+\hbar^2 v_F^2 q^2}\left(\hbar c q+\sqrt{h(q)^2 + \hbar^2 v_F^2 q^2}\right)} = \nonumber \\
h(\Lambda) - \frac{1}{2\pi^2}h(\Lambda)\int\limits_\Lambda^{\Lambda+d\Lambda} \frac{a^2 [\alpha(J)+\alpha(\tilde{J})]d q}{\hbar^2 v_F (c+v_F) q}, \nonumber
\end{gather}
where the additional overall factor of $2$ is due to integration over both positive and negative momenta.
That is, we obtain the corresponding flow equation:
\begin{figure}[t!]
\begin{center}
\includegraphics[width=1\columnwidth]{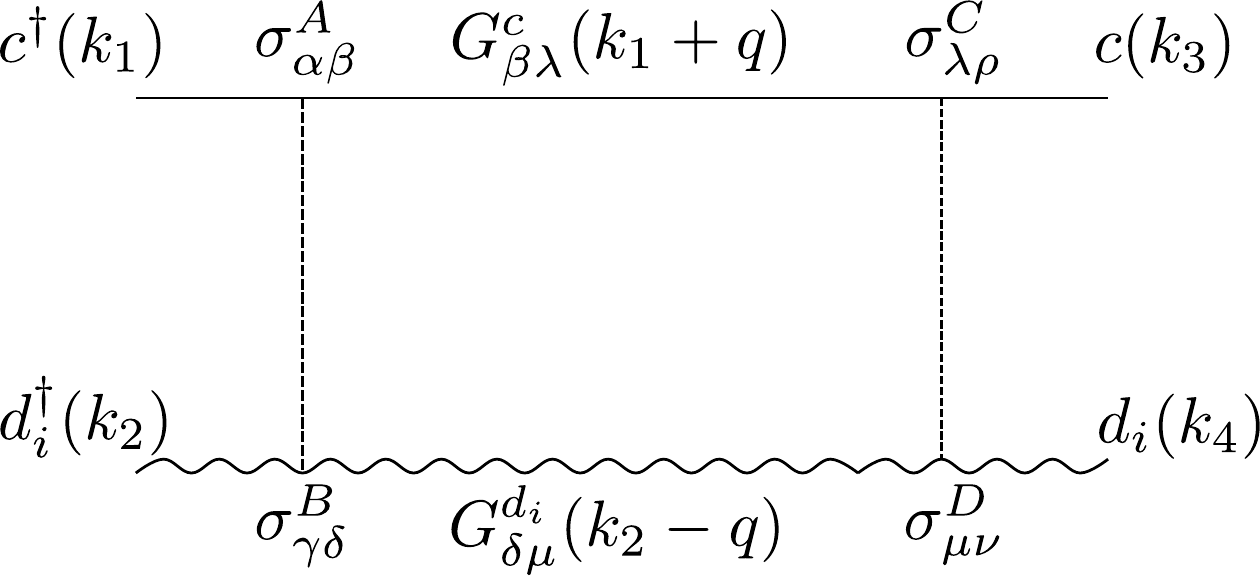}
\end{center}
\caption{Vertex correction to the coupling matrices $\Gamma^{(1,2)}$. \label{fig:Vertex}}
\end{figure}

\begin{gather}
 \frac{dh}{d\log \Lambda} = -\frac{a^2 h}{2\pi^2\hbar^2 v_F(c+v_F)}\left[ \alpha(J) + \alpha(\tilde{J})\right], \label{eq:hRenorm}
\end{gather}
If we ignore for a moment renormalization of other parameters of the model, we can readily conclude:
\begin{equation}
h(\Lambda)=h_0 \cdot \left( \frac{\Lambda}{\Lambda_{UV}}\right)^\gamma \label{eq:h_irrelevance},
\end{equation}
where for further convenience we introduce a notation for the exponent, as it serves as a good measure of the ``irrelevance'' of the process:
\begin{equation}
\gamma= -\frac{a^2}{2\pi^2\hbar^2 v_F(c+v_F)}\left[\alpha(J) + \alpha(\tilde{J})\right],
\end{equation}
If only spin-spin interactions are present
\begin{equation}
\alpha(J)+\alpha(\tilde{J})=-J^2_{zz}-\tilde{J}^2_{zz},
\end{equation}
and the mode mixing is clearly irrelevant in the infrared limit $\Lambda \rightarrow 0$ ($\gamma >0$).

However, this naive treatment is incomplete; to obtain a full picture of interaction effects in this model also requires taking into account renormalization of the coupling matrices $\Gamma^{(1,2)}$, and the Fermi-velocities $v_F$ and $c$.

Renormalization of the couplings is given by one-loop vertex diagram shown in Fig. \ref{fig:Vertex}. The corresponding
momentum integral is also logarithmically divergent, and, omitting the intermediate steps similar to what we have done when computed the backscattering amplitude renormalization, we arrive at the following system of RG flow equations:
\begin{widetext}
\begin{align}
 \begin{array}{cc} \frac{dJ}{d\log\Lambda} = \frac{a}{\pi\hbar(c+v_F)}J\left(J_{00}+J_{0z}-J_{z0}-J_{zz}\right) & \frac{d\tilde{J}}{d\log\Lambda} = \frac{a}{\pi\hbar(c+v_F)}\tilde{J}\left(\tilde{J}_{00}-\tilde{J}_{0z}+\tilde{J}_{z0}-\tilde{J}_{zz}\right) \\ \frac{dJ_{00}}{d\log\Lambda} = \frac{a}{2\pi\hbar(c+v_F)}\left(2J^2+(J_{00} - J_{z0})^2 + (J_{0z}-J_{zz})^2 \right) &  \frac{d\tilde{J}_{00}}{d\log\Lambda} = \frac{a}{2\pi\hbar(c+v_F)}\left(2\tilde{J}^2+(\tilde{J}_{00} + \tilde{J}_{z0})^2 + (\tilde{J}_{0z}+\tilde{J}_{zz})^2 \right) \\  \frac{dJ_{0z}}{d\log\Lambda} = -\frac{a}{\pi\hbar(c+v_F)}\left(J^2 - (J_{00}-J_{z0})(J_{0z}-J_{zz}) \right) & \frac{d\tilde{J}_{0z}}{d\log\Lambda} = \frac{a}{\pi\hbar(c+v_F)}\left(\tilde{J}^2 + (\tilde{J}_{00}+\tilde{J}_{z0})(\tilde{J}_{0z}+\tilde{J}_{zz}) \right) \\ \frac{dJ_{z0}}{d\log\Lambda} = \frac{a}{2\pi\hbar(c+v_F)}\left(2J^2-(J_{00} - J_{z0})^2 - (J_{0z}-J_{zz})^2 \right) &  \frac{d\tilde{J}_{z0}}{d\log\Lambda} = \frac{a}{2\pi\hbar(c+v_F)}\left(-2\tilde{J}^2+(\tilde{J}_{00} + \tilde{J}_{z0})^2 + (\tilde{J}_{0z}+\tilde{J}_{zz})^2 \right) \\  \frac{dJ_{zz}}{d\log\Lambda} = -\frac{a}{\pi\hbar(c+v_F)}\left(J^2 + (J_{00}-J_{z0})(J_{0z}-J_{zz}) \right) &  \frac{d\tilde{J}_{zz}}{d\log\Lambda} = \frac{a}{\pi\hbar(c+v_F)}\left(-\tilde{J}^2 + (\tilde{J}_{00}+\tilde{J}_{z0})(\tilde{J}_{0z}+J_{zz}) \right)\end{array}
\label{eq:JRenorm} \end{align}
\end{widetext}

Fermi velocity renormalization comes from the diagonal part of the self-energy diagram Fig. \ref{fig:SelfEnergy}. Formally speaking, there are two different Fermi-velocities for the two edge chiral modes that renormalize independently:
\begin{gather}
 \frac{dv_F}{d\log\Lambda} = -\frac{4a^2}{\pi^2\hbar^2 (c+v_F)^2}v_F J^2, \label{eq:vfRenorm}\\
 \frac{d\tilde{v}_F}{d\log\Lambda} = -\frac{4a^2}{\pi^2\hbar^2 (c+\tilde{v}_F)^2}\tilde{v}_F \tilde{J}^2, \nonumber
\end{gather}
but we can consistently assume symmetry between them, and impose $J=\tilde{J}$, $v_F=\tilde{v}_F$ at all scales.

In principle, we also have to derive the renormalization group flow for the Fermi velocity $c$, but since 
$v_F \gg c$ in the cases of interest (when the discussed renormalization of backscattering amplitude is strong), and they appear in $1/(v_F+c)$ combination, renormalization of the bath Fermi velocity can be neglected.

Now we are ready to solve flow equations \eqref{eq:hRenorm}, \eqref{eq:JRenorm}, \eqref{eq:vfRenorm} numerically in different regimes, and identify how the backscattering of chiral modes is affected by the environment. To make numerical estimates, we need to agree on the values of bare physical quantities. Fermi-velocity of the edge degrees of freedom in two-dimensional $\mbox{Bi}_2 \mbox{Te}_3$ topological insulators is measured to be 
$v_F \simeq 5\cdot 10^7 \mbox{cm/s}$ \cite{Science-vF}. The spin bath velocity $c$ is a free parameter that can be tuned to any value by choosing a proper environment material, and we find the effect of backscattering suppression to be stronger when $c$ is small, $\sim 10^7 \mbox{cm/s}$, i.e. when the bath is insulating. The in-plane lattice constant for $\mbox{Bi}_2 \mbox{Te}_3$ is
$a = 6.67 \mbox{\normalfont\AA}$. It is interesting to study the model in different regimes and analyze both the role of the spin-spin and the Coulomb interactions, and their interplay.

Thus, we will take the bare backscattering amplitude $ h = 0.1 \mbox{eV}$, and focus on three different cases:
\begin{itemize}
\item{The Coulomb interaction is dominant:
\begin{gather}
 J_{00} = \tilde{J}_{00} = 0.2 \mbox{eV},\\
  J = \tilde{J} = J_{zz} = \tilde{J}_{zz} = 0.
\end{gather}
The energy gap in $Bi_2 Te_3$ is $\Delta E \simeq 0.34 \mbox{eV}$, so we do not want the exchange interactions to be larger than that.}
\item{Spin-spin channel is dominant:
\begin{gather}
   J = \tilde{J} = J_{zz} = \tilde{J}_{zz} = 0.2 \mbox{eV},\\
J_{00} = \tilde{J}_{00} = 0.
\end{gather}
While this case seems quite special since normally the Coulomb interactions are stronger than the $s-d$ exchange, it is instructive to consider this regime as it shows a possibility to use the environment to suppress backscattering and enhance protection of the chiral edge modes.
}
\item{Spin and Coulomb interactions are comparable:
\begin{gather}
 J = \tilde{J} = J_{zz} = \tilde{J}_{zz} = J_{00} = \tilde{J}_{00} = 0.2 \mbox{eV}
\end{gather}
This case appears to be the most non-trivial one as we will see below.
}
\end{itemize}

\begin{figure}[t]
\begin{center}
\includegraphics[width=1\columnwidth]{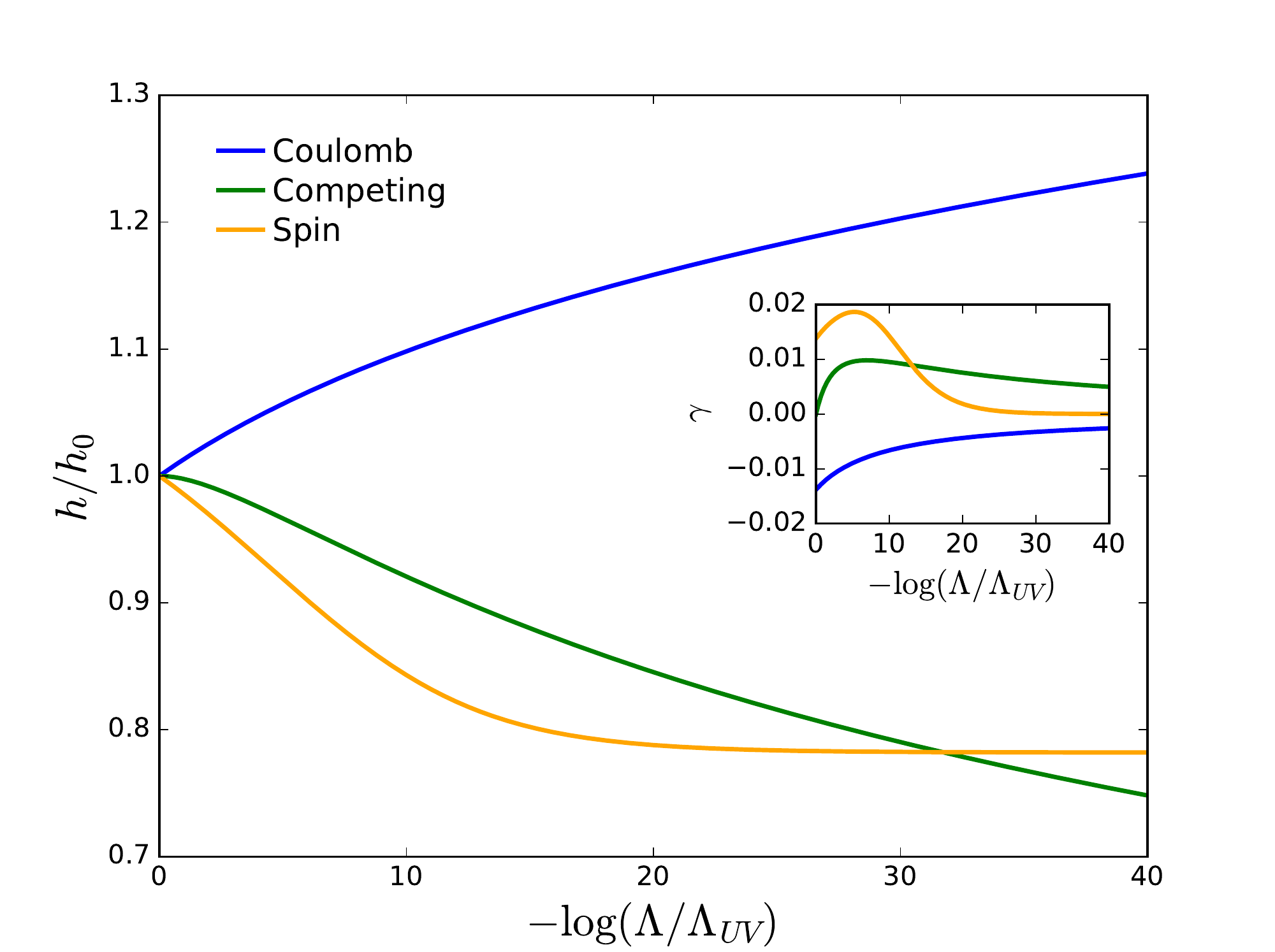}
\end{center}
\caption{RG flows of the backscattering amplitude. The blue curve depicts the Coulomb-interaction dominated case, the yellow one - the case of dominant spin-spin interaction channel, and the green one - the regime of interplay. Inset: the corresponding flows of the ``irrelevance'' parameter $\gamma$.}\label{fig:h_flow_3_cases}
\end{figure}

The numerical solution to the systems of the renormalization group equations in the three mentioned regimes is shown in Fig. \ref{fig:h_flow_3_cases}. One can see that while the pure Coulomb interaction causes enhancement of backscattering, its interplay with the spin-spin coupling and the other induced interaction channels is highly non-trivial. At intermediate energies, if the two competing channels are present, Coulomb reduces the effect of suppressing. However, if one goes to lower energies, it assists the spin interactions in suppressing the process of backscattering, and makes $h$ flowing to zero even when capacity of the spin channel is exhausted, and renormalization of $h$ stopped. 
Another way to see this is to look at the inset of Fig. \ref{fig:h_flow_3_cases}, where renormalization of the exponent of \eqref{eq:h_irrelevance} is shown. Though Coulomb interaction decreases the initial value of the ``irrelevance'' exponent $\gamma$, deep in the infrared it prevents $\gamma$ from flowing to zero. The difference between the two regimes looks rather mild, but since the coupling constants flow towards strong coupling in the infrared, the leading order perturbative analysis tends to underestimate renormalization of $\gamma$, and a stronger effect can be expected.

In this paper, by deriving the leading order perturbative renormalization group flow equations, we have studied how interactions with environment affect the backscattering of chiral modes in helical edge channels. We have discovered that the interplay of the Coulomb and spin-spin interactions between the modes and the environment leads to a non-trivial phase diagram. Dominance of the Coulomb interaction expectedly leads to amplification of the backscattering, making chirality of the propagating modes poorly defined. If only the spin-spin interaction channel is present, the backscattering gets marginally suppressed along the RG flow, receiving a finite negative correction to its bare amplitude. The most interesting situation is when both the interaction channels are at work. Then the Coulomb interaction assists the spin-spin one in suppressing backscattering, making it rather relevant than marginal. The conducted analysis allows us to conclude that the external bath of itinerant spins can be not only dangerous for the chirality of modes in the channel, but also, in certain regimes, can serve as a stabilizer and alleviate the destructive effect of backscattering, restoring the protection of the chiral modes. For a particular example of archetypical 2D topological insulator $\text{Bi}_2\text{Te}_3$, we have estimated that interactions with environment can reduce the backscattering amplitude to $\sim 75\%$ of its original value within a physically reasonable range of energy scales. In systems with smaller Fermi velocities, like InAs/GaSb quantum wells studied among other structures in \cite{Klinovaja1,Klinovaja2} ($v_F=4.6\cdot10^6\mbox{cm/s}$, $a=6.1 \mbox{\normalfont\AA}$), the suppression is even more pronounced, and the backscattering amplitude can be reduced by a factor of 5 or more. One should also keep in mind that the leading order perturbative expansion might provide only the lower bound on the strength of the effect, as the deep IR limit of the model is strongly coupled, and a stronger suppression is expected in the non-perturbative domain, making the studied mechanism a good candidate for protecting chiral modes in edge channels.

\vskip 20pt
We are thankful to Evgeny Stepanov for useful disucussions. The work was supported by the ERC Advanced Grant 338957 FEMTO/NANO and by the NWO via the Spinoza Prize.

\end{document}